\documentclass[conference]{IEEEtran}

\usepackage{graphicx}
\graphicspath{{Figures/}}
\usepackage{subcaption}

\def \sys {\textit{RadioGrapher}}

\usepackage{cite}

\usepackage{color}

\usepackage{amsmath}

\usepackage{amsfonts}

\usepackage{tabularx}

\ifCLASSINFOpdf
\else
\fi

\begin{document}
%

\title{Zero-Calibration Device-free Localization for the IoT based on Participatory Sensing}

\author{\IEEEauthorblockN{Osama T. Ibrahim}
\IEEEauthorblockA{Wireless Research Center\\
Egypt-Japan Univ. of Sc. \& Tech. (E-JUST)\\
Email: osama.ibrahim@ejust.edu.eg}
\and
\IEEEauthorblockN{Walid Gomaa}
\IEEEauthorblockA{Dep. Comp. Sc. \& Eng.\\
E-JUST and Alex. Univ.\\
Email: walid.gomaa@ejust.edu.eg}
\and
\IEEEauthorblockN{Moustafa Youssef}
\IEEEauthorblockA{Wireless Research Center\\
E-JUST and Alex. Univ.\\
Email: moustafa.youssef@ejust.edu.eg}}


%


\maketitle

\begin{abstract}
	Device-free localization (DFL) is an emerging technology for estimating the position of a human or object that is not equipped with any electronic tag, nor participate actively in the localization process. Similar to device-based localization, the initial phase in DFL is to build the fingerprint database which is usually done manually using site surveying. This process is  tedious, time-consuming, and vulnerable to environmental dynamics. Motivated by the recent advances in the Internet of Things (IoT), this paper introduces \sys{}; a system that automates the process of device-free fingerprint calibration in IoT environments. \sys{} leverages the device-based locations of entities in the area of interest in a crowd-sensing manner, aided with Fresnel zones of the wirelessly connected IoT devices to automatically construct a device-free fingerprint. 
	
	Experimental evaluation of \sys{} in an IoT testbed using multiple entities shows that it can construct DFL fingerprints with high accuracy. Moreover, its median localization accuracy is comparable to that of manual fingerprinting. This comes with no calibration overhead, highlighting the promise of \sys{} as a crowdsourcing device-free fingerprint constructor in IoT environments.
\end{abstract}


%
\IEEEpeerreviewmaketitle

\section{Introduction}

    The concept of device-free localization (DFL)~\cite{challenges} has been proposed as a value-added service for wireless networks. It provides the capability of detecting and tracking human entities without requiring the target to carry any device. The DF system depends on the fact that RF signals are affected by the presence of people and objects in the environment~\cite{Robinson2006}. It operates by observing how the human disturbs the radio signal pattern. The physical quantities of the network, in particular the received signal strength (RSS) of a wireless transmitter (AP) at one or more monitoring points (MP) (e.g., laptops, cell phones, other APs, or any IoT device), are processed to detect the existence of a human and estimate her location.
	
	Device free localization technologies are useful in applications where the people being tracked are expected not to cooperate with the system. This may be the case if the human target is deliberately trying to evade the system, such as intrusion detection and burglary prevention applications. This is also the case of physically unable targets, such as helping lonely elders or disabled individuals in an emergency like fire and fall. Moreover, the DFL system does not need any special hardware, as a result, it can use the WiFi infrastructure already installed for data transmission during the day for low cost surveillance during the night; without using any extra hardware.  The IoT environments have the advantage of providing many wirelessly-connected nodes with the existence of abundant smart devices in the area of interest. This gives the DFL system many RSS data streams that potentially cover the whole area without extra hardware deployment. By processing these information, more accurate localization results can be delivered.

	Typically, the DFL process operates in two phases \cite{challenges,Nuzzer2013, PC-DfP2012, DFLAR2017}
	: (a) an offline calibration phase, during which the device-free \emph{fingerprint} of the RSS received from the APs at MPs is constructed for different locations in the area of interest; and (b) an online phase, in which intrusion detection and target tracking are performed by comparing the current RSS readings with the fingerprint records in order to find the best-match result that suggests a location for the target.

	
	The traditional way to construct the DFL fingerprint is by \textit{\textbf{manual} calibration}~\cite{challenges,Nuzzer2013, PC-DfP2012, DFLAR2017}, in which a person travels along all the feasible locations in the area of interest, stands in each location for some time while the MPs record the RSS measurements in the fingerprint associated with this location. 
	This process is tedious, time consuming, and labor intensive. 
	Moreover, every change in the environment leads to changing the RF propagation environment, requiring the same process to be repeated to update the fingerprint. 

	Some systems \textit{diminishes the calibration effort} by manually constructing the fingerprint for a single entity and mathematically duplicate its effect to provide localization for multiple-entities~\cite{SCPL2013,ACE}.
    Those systems can achieve reasonable localization accuracy. However, they still need some manual calibration. 
    

    Other systems eliminate the calibration effort completely using \textit{RF propagation models} to simulate the fingerprinting process~\cite{AROMA2011,RTI2010}. These models predict the WiFi links RSS values with the existence of an entity or more in certain locations, and thereby present an approximated version of the fingerprint. However, this reduction in fingerprint construction overhead comes at the cost of reduced accuracy. 
	
	In this paper, we introduce \sys{}: as a calibration-free DFL system that uses \emph{crowd-sourcing} to \textbf{\textit{automatically}} construct the DFL fingerprint. In particular, regular users of the area of interest during their daily life \textbf{passively/implicitly} participate in constructing the device-free fingerprint, removing the need for the explicit manual labour-intensive fingerprinting process. 
	Moreover, \sys{} also has the advantage of always keeping the fingerprint up-to-date to meet any changes in the environment.
	

	The basic idea of \sys{} is to leverage the \textbf{device-based} localization systems that are already deployed in the area of interest to provide, e.g., navigation services during the day as a way to tag the location of persons and associate them with the \textbf{device-free} fingerprint that will be used in device-free surveillance at night. 
	To achieve this, one main challenge to address is the mismatch between the detected number of persons through the device-based system and the actual number of persons in the environment, since not all persons may be using the device-based system. This can significantly affect the quality of the crowd-sourced fingerprint. 
	\sys{} overcomes this challenge by proposing a calibration-free \textbf{device-free human detector} based on the Fresnel zones of the different IoT devices installed in the area of interest that reveals the mismatching entities, detecting outliers, and increasing the system accuracy. 

    
    Evaluation of \sys{} using two human entities in a smart home over a week period shows that the crowd-sourced fingerprints are $ 98.7\% $ precise,  achieving a median localization accuracy comparable to that of manual calibration. This high accuracy comes with no calibration overhead. 
	
	The rest of this paper is organized as follows. Section~\ref{secOverview} provides an overview on how \sys{} works. Section~\ref{secDetails} gives the details of its core modules and how the device-free detection functionality is delivered. We evaluate the system performance in Section~\ref{secEvaluation}.  Finally, Section~\ref{secConclusion} concludes the paper and discusses future directions.

\section{System Overview\label{secOverview}}
    \begin{figure}
        \centering
        \includegraphics[width=0.9\columnwidth]{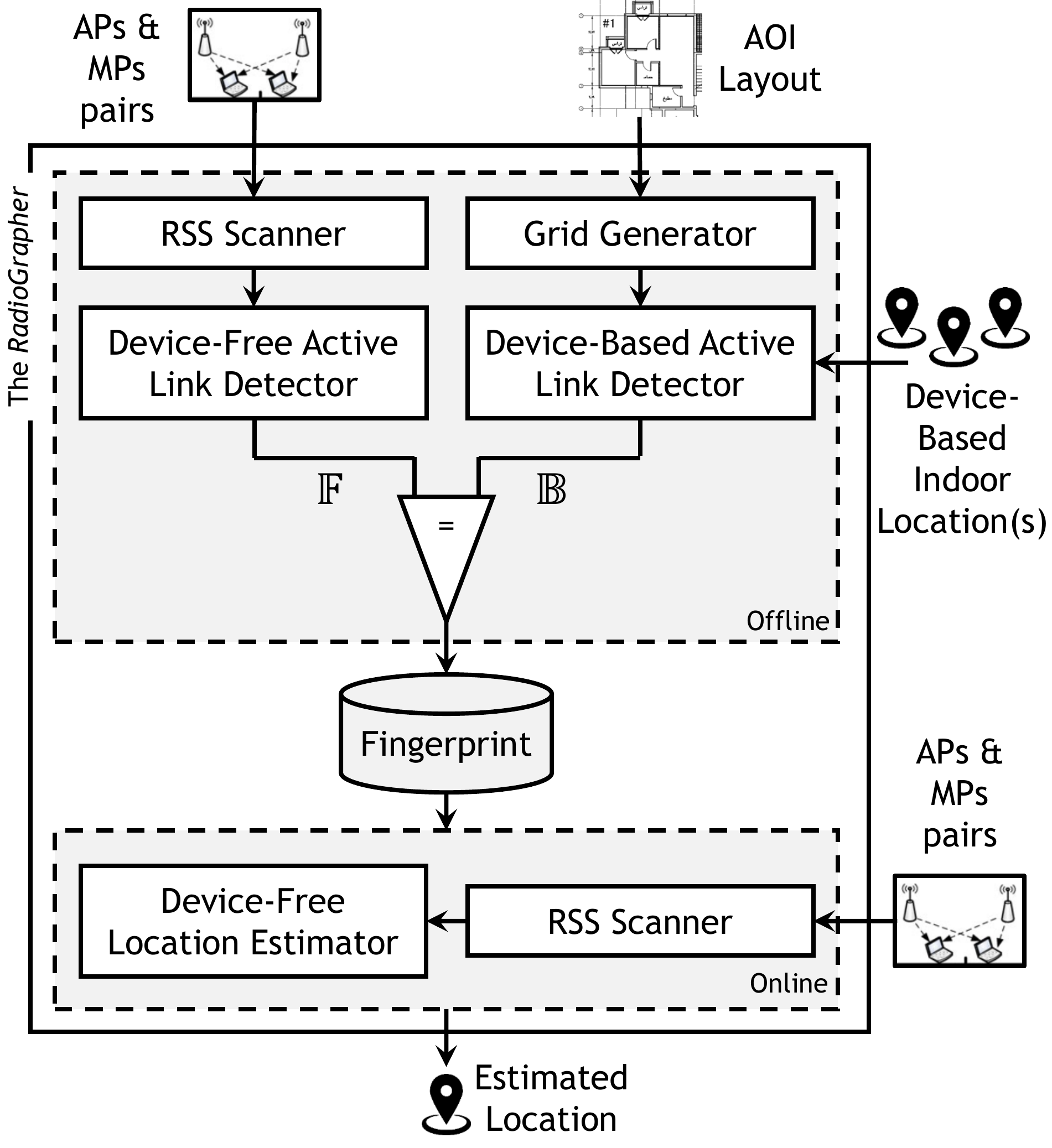}
        \caption{The \textit{RadioGrapher} system architecture.}
        \label{figSystemArchitecture}
    \end{figure}

	The \textit{RadioGrapher} system architecture is depicted in Fig.~\ref{figSystemArchitecture} indicating data flow and the main building blocks. \textit{RadioGrapher} is designed to be deployed in a typical IoT environment, where WiFi technology is already installed for coverage purpose, multiple WiFi-connected IoT devices are present (e.g. smart TVs, sensors, etc), and a \textbf{device-based} localization system is used for localization (e.g. for navigation and/or asset tracking). \sys{} can work with any device-based location tracking technology, e.g. \cite{CrowdInside2012,IncVorr}. Without loss of generality, we use the calibration-free IncVor system in this paper \cite{IncVorr}.
	
	The environment typical users (e.g., smart home residents or company employees), whom we call \textbf{host users}, are incentivized to install a front-end software on their mobile devices to provide them with the IoT connectivity and control as well as the \textbf{device-based} localization functionality such as navigation.

	

	\textit{RadioGrapher} works in two phases: the offline calibration phase and the online tracking phase. In the offline phase, \textit{RadioGrapher} \textbf{automatically} constructs the device-free fingerprint by crowd-sourcing the device-based locations from the daily users of the area of interest. To do that, 
		the \textbf{RSS Scanner} running on the IoT devices reads the pairwise (AP-MP) RSS readings and sends them to a central server for processing. In parallel, the time-stamped \textbf{device-based} locations of the current host users are also transparently sent from the user devices to the server. The server implements and runs the remaining core blocks of the \textit{Radiographer}. 

The \textbf{Grid Generator} module divides the input floor-plan into square cells of arbitrary size, where the center of each cell is a candidate location that requires constructing a fingerprint. If the location of each person in the area of interest is known, the fingerprint construction process would be straightforward. However, since not all users may be running the system front-end software, there may be some users, i.e. ``guest users'' \footnote{We use the term ``host user'' to refer to users who have a device running our front-end software and hence their device-based location is known. ``Guest users'', on the other hand, are those users who may exist in the area of interest for sometime but their locations are not known to the system.}, who are present in the area of interest but their existence and locations are not known to the system. If these guest users are not accounted for, the constructed device-free fingerprint would be erroneous in terms of the locations of the person and their count.

	The remaining part in the \textit{RadioGrapher} offline stage addresses this challenge by detecting the guest users existence. The RSS data streams are analyzed and, based on the existence of guest users, the \textit{RadioGrapher} decides whether to store the fingerprint record or not. 
	Specifically, the \textbf{Device-Free Active Link Detector} declares the active WiFi links that are currently experiencing ``significant'' change in RSS. To differentiate between the active links due to the host and guest users, the \textbf{Device-based Active Link Detector} calculates the links that should be active given the host users locations, and the \textbf{Comparator} module matches them. If there is a mismatch between the two numbers, this reflects the existence of guest users whose device-based locations are unknown. In this case, the \textit{RadioGrapher} discards this fingerprint record and continues to search the data streams for an accurate record. Note that since the system is running all the time and is passive to the users, it is OK to discard records while looking for periods where all the users in the area of interest are host users.  

    During the online phase, the \textbf{Location Estimator} processes the signal strength test vector collected online and runs any traditional device-free localization algorithm, e.g. \cite{Nuzzer2013} to estimate the current users locations. 
    The next section gives the details of the system components.
	
	
\section{The RadioGrapher System\label{secDetails}}
    In this section, we present the details of the \textit{RadioGrapher} system architecture for crowdsoucing the DFL fingerprint. 

    \subsection{RSS Scanner}

        The RSS Scanner is a client-server application that collects the pairwise RSS readings at the MPs (IoT devices) 
        from APs and sends them to server for processing.
        
        The client side is a lightweight background service running on all the MPs of different operating systems; it continuously measures the RSS from the overheard APs. To reduce the scanning time and optimize the transmission bandwidth, the RSS Scanner watches only the APs located inside the area of interest. For $ n $ APs, the client side on each MP collects an $ n $-dimensional vector, whose entries represent the signal strength received from all APs. 
        

\subsection{Grid Generator}

	    Gridding the area of interest is a simple task, yet important, for a scalable localization system. It also reduces the overhead of fingerprint as it allows building the fingerprint while the user is continuously moving. 
	    The Grid Generator module processes the input floorplan by dividing it into square cells of arbitrary length. 
	    The input floor-plan can be either acquired from the building CAD information or automatically generated from crowd-sourced data~\cite{CrowdInside2012}.
	    
	    The fingerprint of a given cell is computed by averaging all the RSS values collected inside this cell (based on the tagged device-based location), this releases the requirement that the user stands at a certain location for a certain time while the system collects the RSS readings, which does not fit the crowd-sourced nature of our system. Moreover, gridding the environment generates scalable fingerprint whose size can be arbitrarily reduced by increasing the cell size. However, increasing the cell size degrades the localization accuracy while enhancing computational efficiency. Therefore, the system designer should trade-off the accuracy and the overhead by configuring the grid spacing to an optimal value. 
	    
A grid cell $ G $ is represented by its center location, which is stored along with the RSS readings collected in this cell.
	   		    
    Note that the same approach can be used to construct a probabilistic radiomap rather than the described deterministic one by simply storing the RSS histogram from different scans instead of averaging it from one scan only. This is beyond the scope of this paper.

        

    \subsection{Device-Free Active Link Detector}
        This module analyzes the RSS readings to detect the parts of the environment that are influenced by \textbf{\textit{any}} user existence. The human detection functionality in \textit{RadioGrapher} is based on the radio link behaviour in the existence of a human around. The RF wave undergoes absorption and reflection that affect the RSS. Therefore, \sys{} uses a Fresnel zone model as a calibration-free human detection method. In this section, we introduce a simple background about Fresnel zones, then we verify their accuracy indoors via preliminary experiments. Finally we give the details of the DF detector.
        
        \subsubsection{Background} \label{secFZBackground}
            The human body is $ 60\sim70 \% $ composed of water, and radio waves have the characteristic of being absorbed by water~\cite{HumanRespiration2016}. Accordingly, if an object is close to the radio link, the RSS through this link will change perceptibly.

            To study the area around a radio link and the reflections due to human object existence within it, a Fresnel zones model was introduced to analyze RF propagation. 
            Since then, it has been applied in outdoor localization~\cite{FZoutdoor2016} and human respiration detection~\cite{HumanRespiration2016}, among others. In this context, the Fresnel zone of a radio link is a set of concentric ellipsoids with foci in the pair of (transmitter-receiver) nodes connecting this link. Each zone describes a constructive or destructive interference between the direct wave (transmitted over line of sight) and the reflected wave over the human body existing in the zone. The radius of the $ z $th Fresnel zone at any point $ P $ in between the two wireless nodes is given by~\cite{FZoutdoor2016}:
            \begin{equation}
                r_z = \sqrt{\frac{z \lambda d_1 d_2}{d_1+d_2}}, \quad \lambda = \frac{c}{f}
                \label{eqFZPointRadius}
            \end{equation}
            where $ \lambda,c,f $ are the wavelength, speed, and frequency of the radio wave respectively; and $ d_1,d_2 $ are the distances of the point $ P $ to the two nodes. The maximum radius ($ R_z $) is found at the midpoint where $ d_1=d_2=D/2 $, $ D $ is the distance between the two nodes:
            \begin{equation}
                R_z = \sqrt{\frac{z c D}{4 f}}
            \end{equation}
            
        Fresnel zones are used by communication engineers and systems designers to secure a space around the transmission link that is free of obstruction to allow signal transmission with minimum destruction. \sys{} exploits it in a reverse way to detect the human obstacles by probing the interference pattern in the received signal.

        \subsubsection{Preliminary Feasibility Experiments}

\begin{figure}
    \centering
        \centering
        \includegraphics[width=0.6\columnwidth]{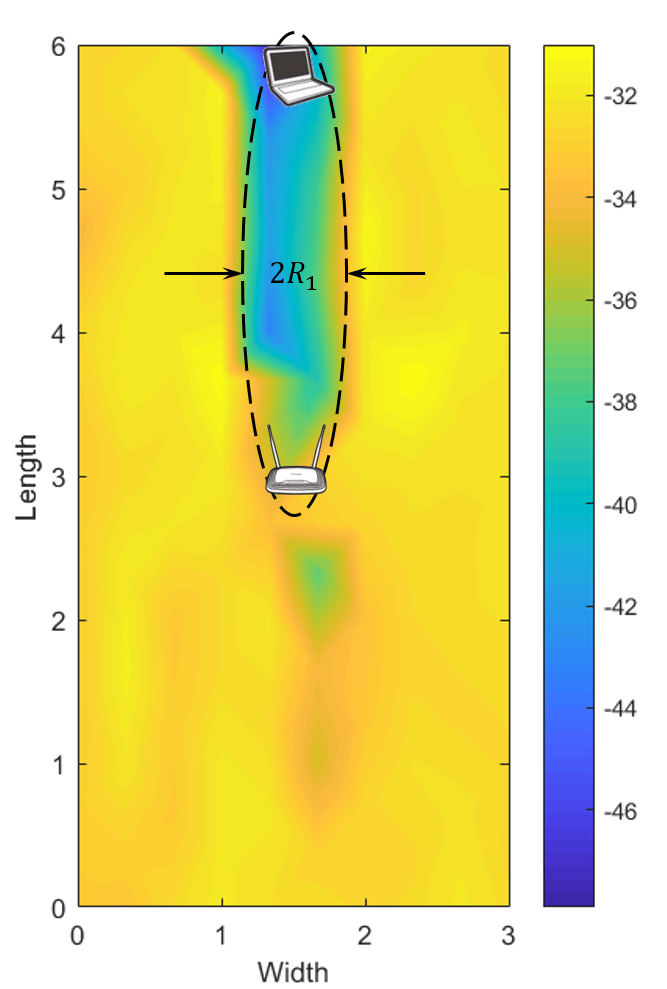}
        \caption{Fresnel Zone of Single-Radio-Link Environment.}
        \label{figSingleLink}
\end{figure}
            To verify the Fresnel zone concept for our system, we implemented a single WiFi link in a clear hall and studied the effect of human existence on the RSS readings. The link nodes operate at $ 2.4 $ GHz frequency with $ 3 $ m intermediate distance. A person stands at each of a uniformly distributed 100 locations in a $ 6 \times 3 $ m$ ^2 $ hall until 300 RSS readings are collected. The average RSS value per location is calculated and its variation against the location coordinates is plotted. Fig.~\ref{figSingleLink} shows the heatmap of this variation.
            Given that more than $ 70\% $ of energy is transferred through the first Fresnel zone~\cite{FZbook2000}, the heatmap in Fig.~\ref{figSingleLink} validates the Fresnel zone model for our application. 
            The results of single link analysis demonstrates the promise of the Fresnel zone model to detect a person indoors. In particular, a simple threshold-based algorithm on the RSS average value can be used for detection of a person inside the first Fresnel zone.
	        

	    \subsubsection{Device-Free Human Detection}
	    
	        The \textit{Device-Free Active Link Detector} module computes the average of each stream within the area of interest and compares it with silence periods (zero-entity) to declare the link as active or not. 
	        The end goal is to detect the total number of users in the area. Based on the Fresnel model, if a person is found in the link zone, the RSS will experience a significant change.
	        
	        As a practical consideration, network interface cards characteristics are different from one MP to another. This leads to a heterogeneity issue in terms of sampling rate and RSS value. Each link $ i $  has its own stream length $ q_i $ and silence average $ \overline{s}^o_i $. The average of each stream is calculated and, to handle the MPs heterogeneity, the \textit{relative} absolute difference is used for thresholding.
	        \begin{equation}
	            \Delta s_i = \left|\frac{\overline{s}_i-\overline{s}^o_i}{\overline{s}^o_i}\right|, \quad \overline{s}_i = \frac{1}{q_i}\sum_{j=1}^{q_i}s_j
	        \end{equation}
	        
	        
	        A link $ i $ is declared active if $ \Delta s_i > \tau $, where $ \tau $ is the device-free activation threshold. 
	       The set of detected device-free active links, $ \mathbb{F} $, is passed to the \textit{Comparator}.
	        We evaluate the effect of $ \tau $ on performance in Section \ref{secEvaluation}.

	\subsection{Device-Based Active Link Detector}
	    The main goal of this module is to determine the data links that should be active when a target ``host user'' is found at a certain radio map location by applying the Fresnel zone model discussed in Section~\ref{secFZBackground}. This is used in combination with the previous section to detect the existence of ``guest users''. Based on the tagged locations of APs and MPs on the input floorplan, Equation~\ref{eqFZPointRadius} draws an ellipse around each WiFi link; a link is declared active if at least an input device-based location is located inside its ellipse. 

	    By definition, the ellipse border draws the locus of a point whose sum of distances to two focal points is constant~\cite{ConicSections}. The constant value is the major axis length, $ 2a $, and the focal points, $ F_1,F_2 $, are represented here by the MP and AP locations. A device-based location $ x_i $ belongs to the Fresnel ellipse if
	    \begin{equation}
	        \lVert x_iF_1 \rVert + \lVert x_iF_2 \rVert \le 2a, \quad a = \sqrt{R_z^2+\left(\frac{D}{2}\right)^2}
	    \label{EqDB}
	    \end{equation}
	    A system parameter, $ z $ , specifies the order of Fresnel zones to be considered in ellipse calculation. The \textit{Device-Based Active Link Detector} searches the input device-based locations for the same timestamp of the currently-processed RSS readings, calculates the enclosing Fresnel zones, and finally sends them as the set of device-based active links, $ \mathbb{B} $, to the \textit{Comparator} to compare them to the device-free active links.
	   

	\subsection{Comparator}
	    The \textit{Comparator} processes the relative absolute 
	    differences of all streams and the DF and DB Active Link Detectors outputs, $ \mathbb{F}, \mathbb{B} $, to detect ``guest users'' existence, i.e. users that are not recognized by the \textbf{device-based} system and hence should be considered as outliers. Guest users should activate more WiFi links than the set of DB active links. The existence of guest users is declared if $ \mathbb{F} \setminus \mathbb{B} \ne \emptyset $. In this case, the current fingerprint record is discarded and the system continues searching the RSS streams for valid entries. Otherwise, the \textit{Comparator} updates the grid cells that contains the estimated device-based locations.

    \subsection{Location Estimator}\label{secOnline}
        The goal of this module is to estimate the user location in the online tracking phase in a device-free manner, i.e. without the users carrying any devices. The problem definition is: given a WiFi signal strength test vector $ S^t $, we want to calculate the users locations by comparing the test vector against all records in the generated fingerprint. The estimated location is the center of the grid cell associated with the most similar fingerprint record to $ S^t $. 
        
        The vectors are compared based on the Euclidean distance in the signal strength space. Specifically, the location estimator returns the fingerprint record that satisfies 
        \begin{equation}
            arg\min_r \sqrt{\sum_{j=1}^k \left ( S^t_j-S^r_j \right )^2}
        \end{equation}
        Where $ k $ is the number of data streams in the environment. The output location is the center of the grid cell $ r $.
        
\section{System Evaluation \label{secEvaluation}}
    In this section, we evaluate the performance of the \textit{RadioGrapher} in a typical IoT environment. We start by describing the experimental testbed and the data collection process, followed by evaluating the effect of system parameters on performance based on certain metrics. Finally, we compare the \textit{RadioGrapher} localization with the traditional manual fingerprinting techniques, \textit{which represent the best accuracy}.

	\subsection{Experimental Testbed and Data Collection}
        \begin{figure}
            \centering
            \includegraphics[width=0.85\columnwidth]{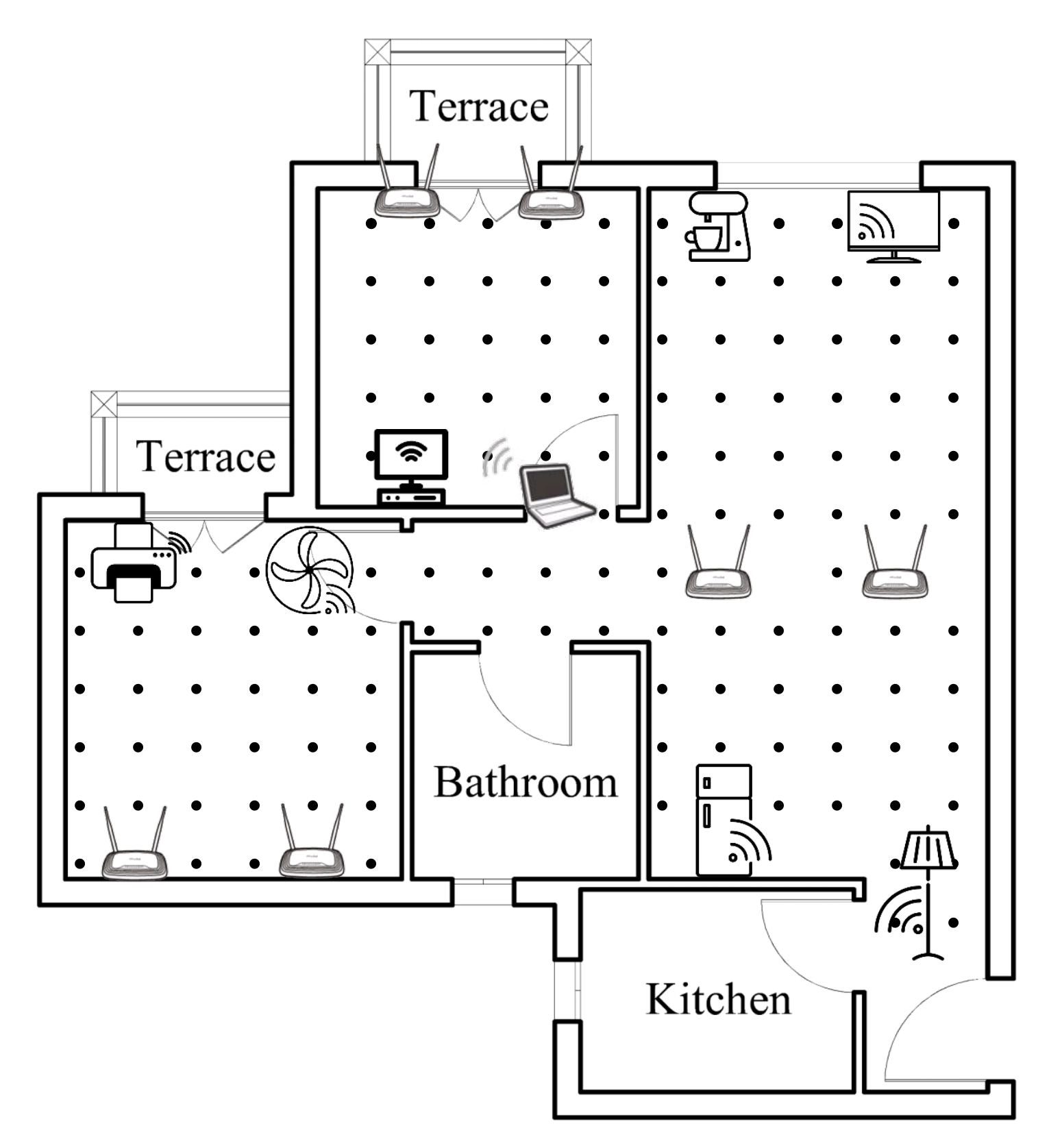}
            \caption{Testbed Layout. Each dot represents a test point.}
            \label{figTestbedLayout}
        \end{figure}

\begin{table}
    	    \begin{tabularx}{\columnwidth}{|X|l|}
    	        \hline
    	        \textbf{Users inside the testbed } & \textbf{No of scenarios}\\
        		\hline
        		One host only & $ 131 $\\
        		\hline
        		A host and a guest in the same zone & $ 8 $\\
        		\hline
        		A host and a guest in different zones within the same room & $ 10 $\\
        		\hline
        		A host and a guest in different rooms & $ 22 $\\
        		\hline
        		Silence, zero users & $ 1 $\\
        		\hline
        	\end{tabularx}
        	\caption{Data collection scenarios. 
            }
        	\label{tableCollectedData}
        \end{table}

        We deployed the \textit{RadioGrapher} in a smart home testbed with the layout shown in Fig.~\ref{figTestbedLayout}. This configuration fits many IoT applications, where multiple home devices are WiFi-enabled. The testbed spans an area of $ 49 $ m$ ^2 $ with $ 2.75 $ m ceiling height. 
        The access points are different models of Netgear N300 operating on the $ 2.4 $ GHz frequency band. 
        
	    To evaluate the \textit{RadioGrapher}, two volunteers of different heights (180, 160 cm) and gender (a male and a female), collected the necessary data for evaluation. This is done by recording the pairwise RSS measurements at the MPs from all the APs in the environment when one or two targets are located at known locations in the area of interest. Test points are uniformly distributed on a grid with a $ 55 $ cm spacing as dotted on Fig.~\ref{figTestbedLayout}.

	    The collected data consists of two datasets, one-user and two-users depending on the number of persons in the environment. The one-user dataset is collected when only one user is located in 131 different locations in the testbed. Whereas, the two-users dataset is collected when they are standing at 20 different location combinations. This results in 40 readings characterizing the RSS behaviour in case of one host and another guest user in the area of interest. Table~\ref{tableCollectedData} details the covered scenarios and their data sizes. 
	    The data was collected over five days, at different times in day and night.
	    
	   
	\subsection{Performance Metrics}
	    
	    To compare the generated fingerprint with the ground-truth manual fingerprint, all readings of the two collected testbeds are fed to the system and finally the stored database is compared to the manual fingerprint ground-truth based on two performance metrics: Precision and Recall. When feeding the system with the collected datasets we are expecting some readings to be stored because they were generated by only host users within the area of interest, and the others are expected to be discarded due to mismatch during the existence of guest users. \textit{Precision} is the fraction of the correctly-stored readings among all the recorded fingerprint, while \textit{Recall} is the fraction of the correctly-stored readings to the total amount of test readings that are expected to be stored.
	    
	    The localization accuracy of \sys{} is compared against other fingerprinting systems in terms of localization error. To test a system, a human target is found at some known location and the error is calculated as the Euclidean distance between the location estimated by the system and the actual location.

\begin{figure*}
    \centering
    \begin{minipage}{0.24\textwidth}
        \centering
        \includegraphics[width=\textwidth]{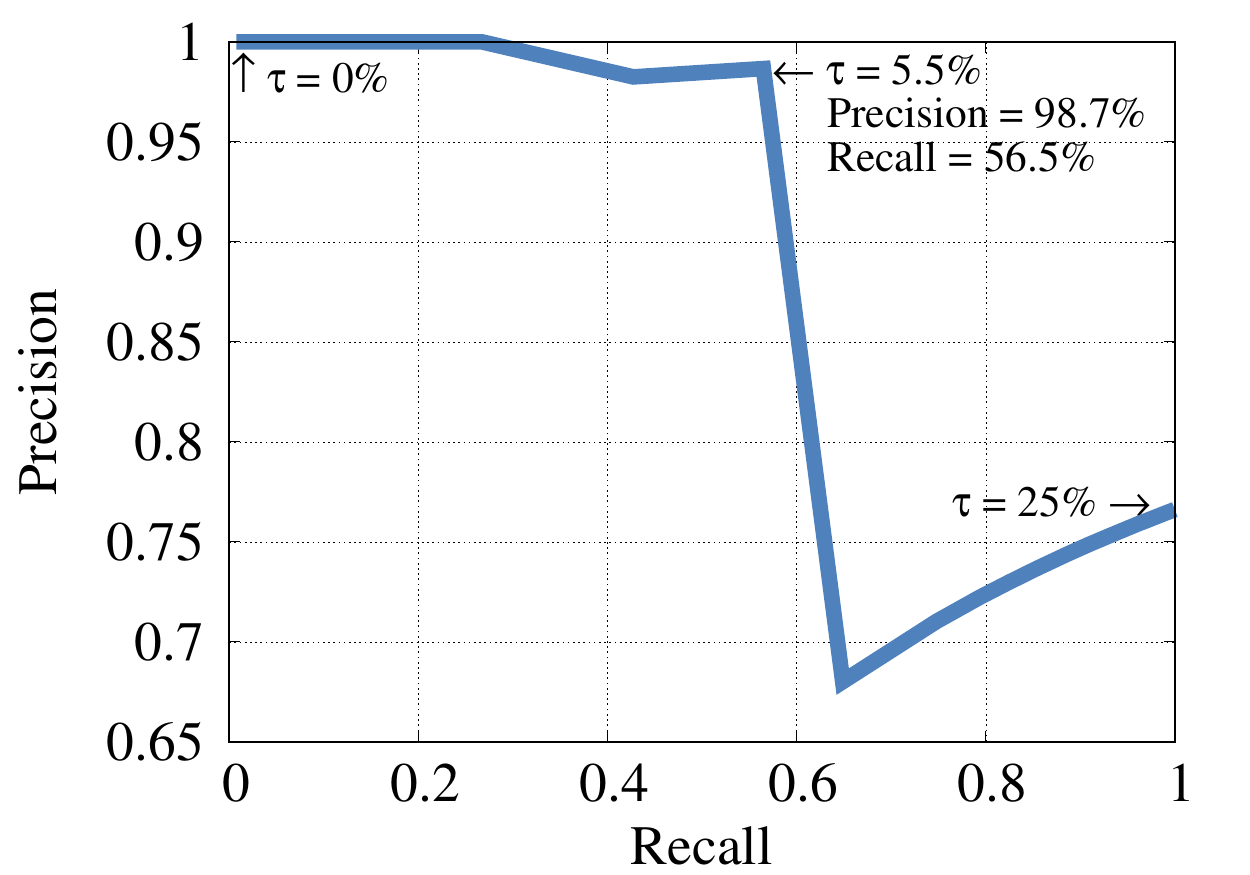}
        \caption{Precision-Recall curve for changing the DFL activation threshold.}
        \label{figEvaluThreshold}
    \end{minipage}
    \hfill
    \begin{minipage}{0.24\textwidth}
        \centering
        \includegraphics[width=\textwidth]{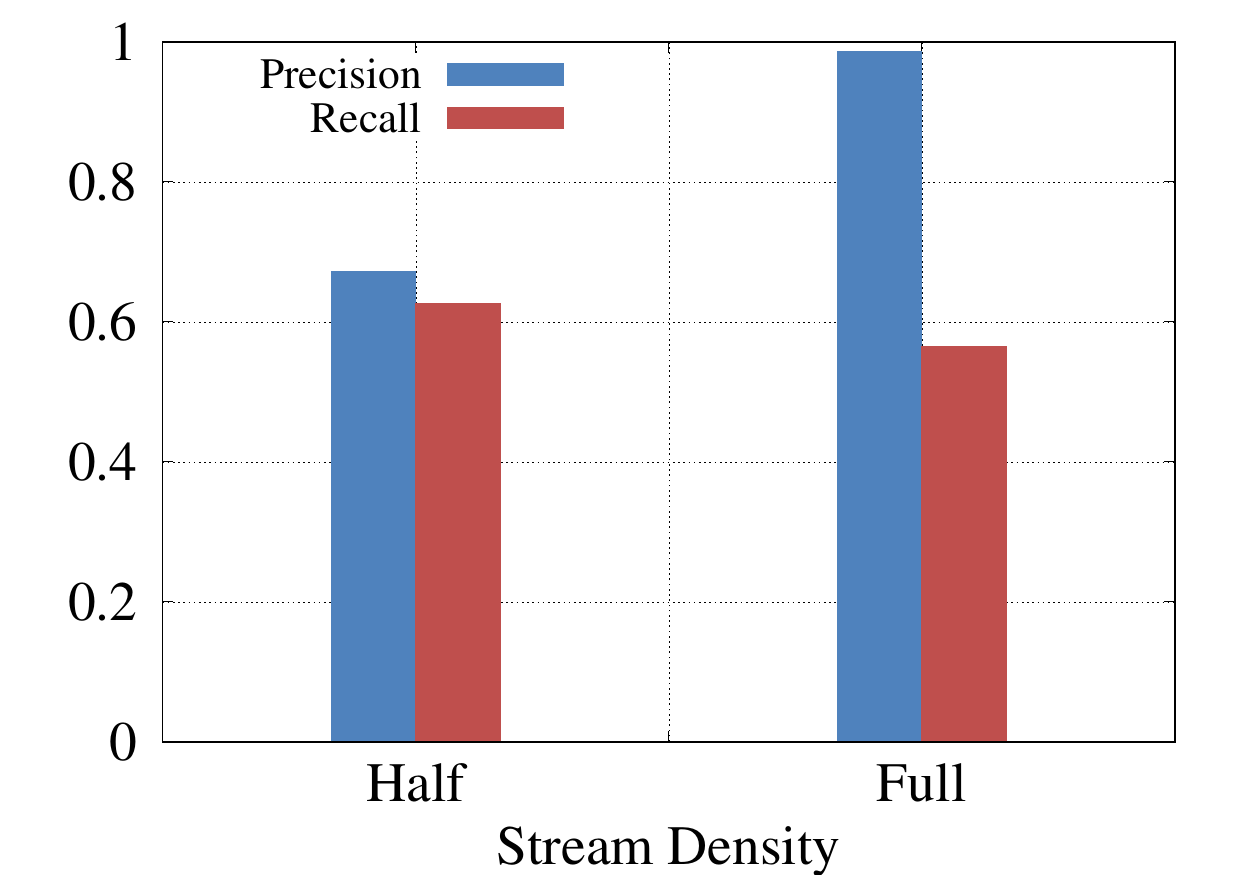}
        \caption{Effect of number of streams per room on the performance metrics.}
        \label{figEvaluStreamCount}
    \end{minipage}
    \hfill
    \begin{minipage}{0.24\textwidth}
        \centering
        \includegraphics[width=\textwidth]{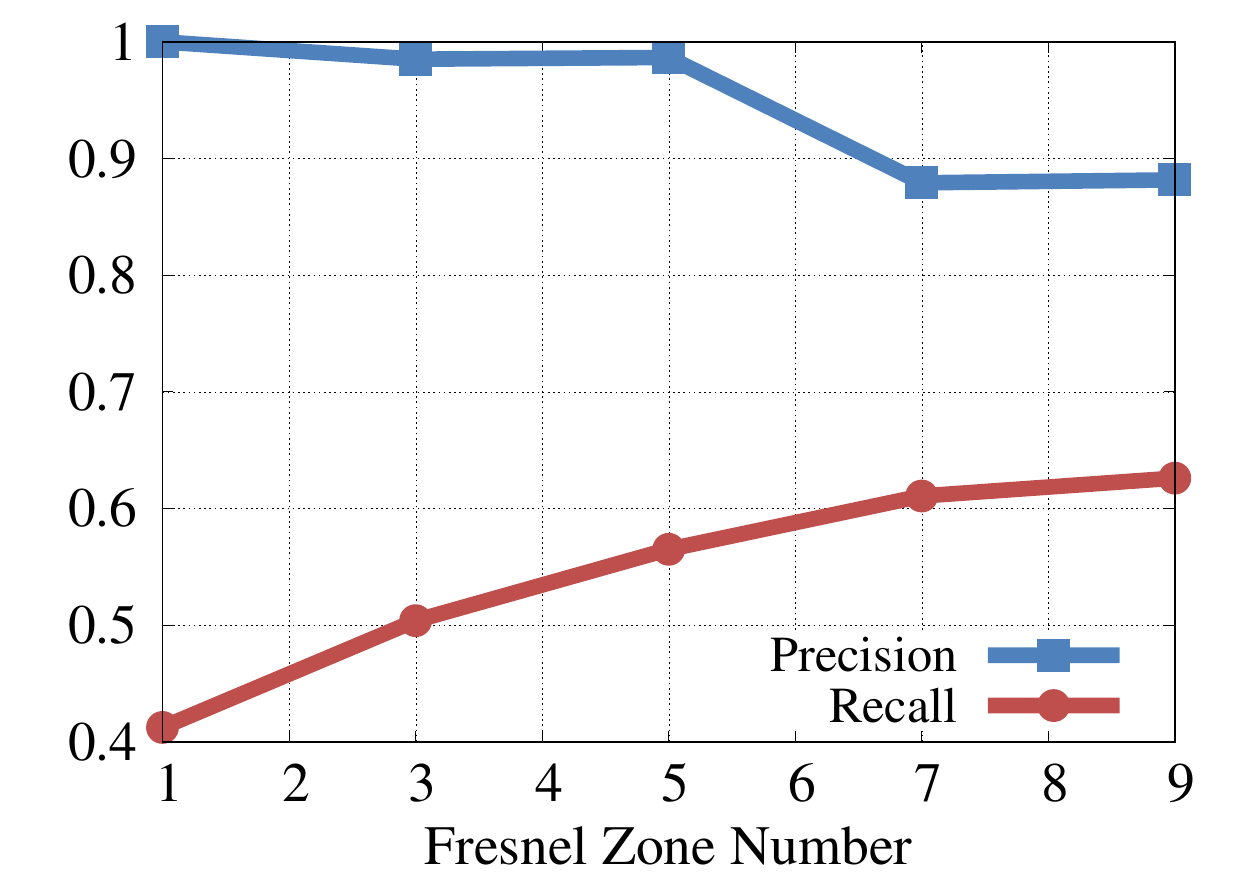}
        \caption{Effect of Fresnel zone order on the performance metrics.}
        \label{figEvaluFresnelOrder}
    \end{minipage}
    \hfill
    \begin{minipage}{0.24\textwidth}
        \centering
        		\includegraphics[width=\textwidth]{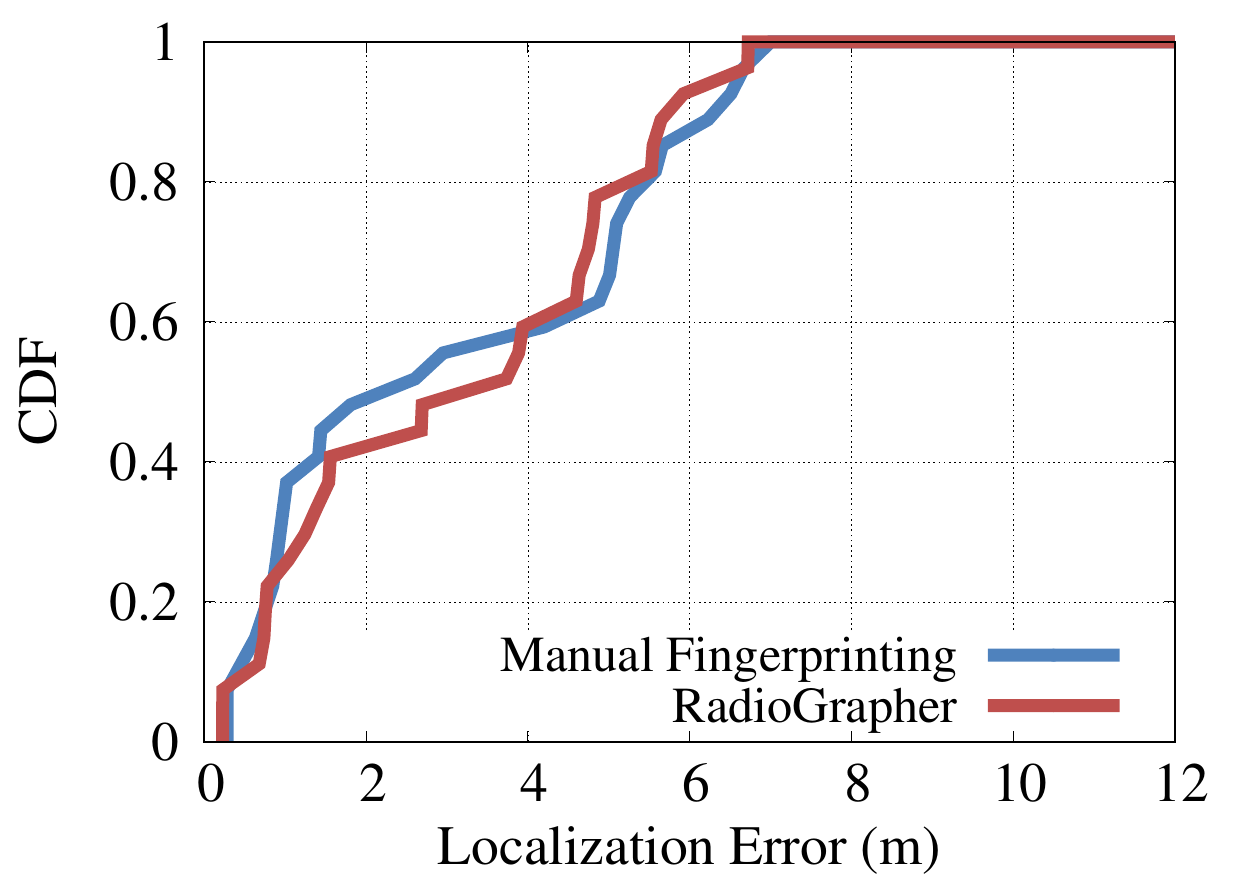}
        		\caption{CDF of localization error of \textit{RadioGrapher} against  manual fingerprinting.}
        		\label{figCDF}
        
    \end{minipage}%
\end{figure*}  
	    
	\subsection{Effect of System Parameters}

	    In this section, we evaluate the effect of changing the system parameters on the generated fingerprint including detection threshold, Fresnel zone order, and number of RSS data streams. 
	    Table~\ref{tableDefSysParam} shows the default parameters values used throughout the evaluation section. 

	    \subsubsection{Effect of DFL activation threshold}
            Figure~\ref{figEvaluThreshold} shows the effect of device-free active link detection threshold on the performance metrics. 
            The links of the set difference $ \mathbb{F} \setminus \mathbb{B} $ can be activated either due to either the existence of guest users (true positive guest detection) as dominant in higher thresholds, or noise and multipathing with no guests (false positive guest detection) as dominant in lower thresholds. 
            Therefore, as shown in Fig.~\ref{figEvaluThreshold}, increasing the threshold form zero to $ 25\% $ reduces the detection of guest users that, in turn, decreases the Precision. This increase in threshold also decreases
            the number of erroneously activated links due to noise and multipath in $ \mathbb{F} \setminus \mathbb{B} $, decreasing the discarded readings and, consecutively, increases the Recall.
            
           Note that in our system, \textbf{ \textit{Precision should have higher weight/importance than Recall}}: whereas lower Recall leads to just losing the opportunity to store some fingerprint records that can be compensated by later data,  lower Precision collects erroneous records in the database, degrading the system performance.

        \subsubsection{Effect of streams density}
            To understand the effect of the number of data streams on  \sys{}, we plot the Precision and Recall with full and half stream density covering the whole testbed  (Fig.~\ref{figEvaluStreamCount}). Any two adjacent pairs of nodes in the testbed generate $ 4 $ data links. The full stream density implements the whole $ 4 $ links while the half density utilizes $ 2 $ of them only in the system calculations. 
            The figure shows that, reducing the number of data streams decreases the size of difference set $ \mathbb{F} \setminus \mathbb{B} $ which minimizes the chance of a guest entity to be detected, this increases the number of erroneously recorded readings in the fingerprint leading to lower Precision. On the contrary, this reduction in difference set also reduces the false positives of zero-guest cases to be detected as guest; then the discarded readings diminishes leading to higher Recall!! 
            As mentioned before, we seek for higher Precision, so we use the full stream density as the default value for this parameter despite the slight decrease in the Recall. 
            It is still few streams that achieve high accuracy.
            
        \subsubsection{Effect of the Fresnel zone order}

           Figure~\ref{figEvaluFresnelOrder} shows the system performance with widening the zone used in device-based active link detection, i.e. the parameter $ z $ in Eq.~\ref{EqDB}. Evident from the figure, with higher order zones, the Precision is decreased and Recall is increased. This can be explained by noting that increasing the Fresnel zone order activates more links in the device-based detector at the same device-based locations which, from one hand, allows some guest users to be undetected as they are covered by the same device-based active zones of host users. This leads to storing the fingerprint records in these cases resulting in lower Precision. 
         
           On the other hand, the existence of a host user in a zone may activate some neighboring zones due to the noise and multipath effects, which leads to more neglected readings and lower Recall. With increasing the Fresnel zone order, those neighbors are counted in the device-based active set leading to less neglected readings and higher Recall. We use a Fresnel zone order of five as our default value.
           
           
           
        \subsection{Comparison with Other Systems}
        	
        	We evaluate the localization accuracy of the \sys{} automatically generated fingerprint as compared to traditional manual calibration as the system that gives the best accuracy. Both techniques employ the same estimation algorithm in the online phase (Section~\ref{secOnline}). This allows us to evaluate the quality of the automatically generated fingerprint under the same condition. Comparison is done through an independent dataset of $ 27 $ readings collected at uniformly distributed random locations in the testbed area. Figure~\ref{figCDF} compares the localization error CDF of both techniques. 
        	The figure illustrates that \textit{RadioGrapher} has comparable accuracy to that of manual fingerprinting with identical percentiles starting from $ 60\% $ and a slight decrease of median accuracy of $ 1 $ meter. This comes with no calibration overhead.

\begin{table}
    	    \begin{tabularx}{\columnwidth}{|X|l|l|}
    	        \hline
    	        \textbf{Parameter} & \textbf{Range} & \textbf{Default value}\\
        		\hline
        		DFL Activation Threshold ($ \tau $) & $ 0-25\% $ & $ 5.5\% $\\
        		\hline
        		Fresnel zone order ($ z $) & $ 1-9 $ & $ 5 $\\
        		\hline
        		Number of data streams per room & $ 2-4 $ streams & $ 4 $\\
        		\hline
        	\end{tabularx}
        	\caption{Default System Parameters}
        	\label{tableDefSysParam}
        \end{table}

\section{Conclusion} \label{secConclusion}

    We presented the design, implementation, and evaluation of the \textit{RadioGrapher} as a system that automates the device-free fingerprint construction process by crowd-sourcing from device-based locations in IoT environments. 
    As part of \textit{RadioGrapher}, we introduced a
    novel, low-overhead calibration-free Fresnel-zone-based device-free detection algorithm by which \textit{RadioGrapher} checks the existence of unattached entities, and accordingly approves storing or discarding the fingerprint records.
    
    Evaluation of \textit{RadioGrapher} in an IoT-enabled indoor environment with multiple human entities shows that the fingerprint crowd-sourced by  \textit{RadioGrapher} are almost $ 100\% $ precise. This leads to a localization accuracy comparable to manual fingerprinting under the same deployment conditions, without the tedious, labor intensive, and time-consuming calibration phase. 
    
    Currently, we are expanding \sys{} in multiple directions including reducing the required number of streams, extending \sys{} to other device-free applications, among others.




\bibliographystyle{IEEEtran}
\bibliography{references.bib}

\end{document}